\documentclass[a4paper,fleqn,usenatbib]{mnras}

\usepackage[T1]{fontenc}
\usepackage{ae,aecompl}

\usepackage{graphicx}	
\usepackage{amsmath}	
\usepackage{amssymb}	

\title[Spectroscopic modelling of HMXBs]{Spectroscopic modelling of two high-mass X-ray binaries, Cyg X--3 and 4U 1538--522}

\author[Shaw and Bhattacharyya]{
Gargi Shaw$^{1}$\thanks{E-mail: gargishaw@gmail.com},
Sudip Bhattacharyya$^{1}$
\\
$^{1}$Department of Astronomy and Astrophysics, Tata Institute of Fundamental Research,
 1 Homi Bhabha Road, Colaba,
Mumbai 400005, India\\
}

\date{Accepted ... . Received 2020; in original form 2021}

\pubyear{2015}

\begin{document}
\label{firstpage}
\pagerange{\pageref{firstpage}--\pageref{lastpage}}
\maketitle

\begin{abstract}
We report a detailed modelling of soft X-ray
emission lines from two stellar-wind fed Galactic high mass X-ray binary (HMXB) systems, Cyg X-3 and 4U 1538-522, and estimate physical parameters, e.g., hydrogen density, radiation field, chemical abundances, wind velocity, etc. The spectral synthesis code CLOUDY is utilized for this modelling.
We model highly ionised X-ray spectral lines 
such as Fe XXV (6.700 keV), Fe XXVI (6.966 keV), and reproduce the observed line flux values. 
We find that for Cyg X--3 and 4U 1538-522, the inner radius of the ionised gas is at a distance of 10$^{12.25}$ cm and 10$^{10.43}$ cm respectively from the primary star, which is the main source of ionisation.
The densities of the ionised gas for Cyg X--3 and 4U 1538--522 are found to be $\sim$ 10$^{11.35}$ cm$^{-3}$ and 10$^{11.99}$ cm$^{-3}$, respectively. The corresponding  wind 
velocities are 2000 km s$^{-1}$ and 1500 km s$^{-1}$. The respective predicted hydrogen
column densities for Cyg X--3 and 4U 1538--522 are $10^{23.2}$ cm$^{-2}$
and 10$^{22.25}$ cm$^{-2}$. In addition, we find that magnetic field affects the strength of the spectral lines 
through cyclotron cooling. Hence, we perform separate model comparisons including magnetic field for both the sources. 
Most of the parameters, except the hydrogen column density, have similar values with and without magnetic field. We estimate 
that the most probable strength of the magnetic field for Cyg X--3 and 4U 1538--522, where the Fe XXV and Fe XXVI lines originate, is $\sim$ 10$^{2.5}$ G.

\end{abstract}




\section{Introduction}
X-ray binary (XRB) systems are co-rotating close binary stellar systems where the compact star is a neutron star or a 
black hole or a 
white dwarf, and the donor/companion star is a normal star or a white dwarf \citep{{Lewin1995},{Sturm2012}}. 
XRB systems are classified in different groups depending on the nature of the compact star and the mass of the companion star \citep{Reig2011}. High-mass 
X-ray
binary (HMXB) systems are one among these classes where the compact star is a neutron star or a black hole, and the companion star is a massive star with 
a mass typically greater than ten solar mass \citep{Liu2006}. 
 HMXB systems emit strong X-ray radiation (luminosity $\sim$ 10$^{36}$ -- 10$^{40.5}$ erg s$^{-1}$) which are powered by mass 
accretion \citep{Zeldovich1966}.
Such accretion occurs via stellar wind \citep{Bondi1944} for most sources, but in some cases also through the Roche 
Lobe overflow \citep{Frank1992}. 

In the process of mass accretion via stellar wind, matter is blown away from the companion star and a small fraction of this matter gets captured by the 
compact object. 
The companion star of HMXBs are very massive 
and hence the matter gets thrown away at a high velocity close to a few thousand km s$^{-1}$. In such a system, 
the compact star is deeply embedded 
in the 
wind and the accreted matter 
heats up and shines in X-rays. Many HMXBs have so far been studied \citep[\& references therein]{{Giacconi1967},{Reynolds1992},{Falanga2015}}, and the X-ray spectral
lines observed from them are sensitive to the underlying physical conditions of both the compact and the companion stars.
Hence, if investigated properly, these X-ray lines can 
reveal a plethora of 
information about the binary systems. 

In this work, we aim to do such a spectroscopic modelling of two HMXB systems using CLOUDY, 
and determine the underlying physical 
conditions like density, radiation field, chemical abundances, etc.
For this purpose, we select two Galactic HMXB systems, 
Cyg X--3 and 
4U 1538--522, which show multiple highly ionised X-ray spectral lines, including Fe XXV (6.700 keV) and Fe XXVI (6.966 keV). 
Fe XXVI and Fe XXV have H-like and He-like iso-electronic sequences, respectively. The He-like Fe XXV
K$\alpha$ complex includes three transitions involving
1s$^2$--1s2p, resulting into w(1S--1P), x(1S--3P2) and y(1S--3P1). The transition involving 1s$^2$--1s2s 
results into the forbidden line z 
(1S--3S) \citep{{Bianchi2002},{Bianchi2004}}. In a photoionised gas, H-like transitions 
are mainly produced by recombination, whereas the He-like transitions are produced by recombination and resonant
scattering. 
In addition to this, the He-like transitions have a wide range of transition probabilities making them density 
sensitive \citep{Osterbrock2006}. 
 
The neutron star in an HMXB has a strong surface magnetic field of $\sim$ 
10$^{12}$ G \citep{Shapiro1984},
which decreases as 
$\mu/r^3$, 
where $\mu$ is magnetic 
dipole moment and $r$ is the distance from the centre of the neutron star. On the other hand, the magnetic fields around black holes are expected to be small. 
For example, \citet{Dallilar2017} 
 measured 
the magnetic field of the black hole V404 to be around a few hundred G. In many HMXBs, the companion star, and hence its wind, can also have a significant magnetic field.
It is challenging to measure the magnetic field of the accreted material in an HMXB using Zeeman splitting or polarisation observations.
Here, we suggest a plausible way to estimate the upper 
limit of the local magnetic field of the irradiated stellar 
wind of HMXBs, as mentioned below.

The presence of a magnetic field helps in cooling a high-temperature
ionised gas through cyclotron emission. It takes dominant part in
thermal cooling and contributes to gas pressure. This cooling is proportional to B$^{2}$. 
Hence magnetic field is capable to change the 
strength of emission/absorption 
lines from a high-temperature ionised gas. Therefore, the strengths of lines originating in such environments can be a suitable way to estimate an upper 
limit of the local magnetic field. 


The HMXB Cyg X--3 was first observed by \citet{Giacconi1967} and 4U 1538--522 was first observed using UHURU 
satellite  by \citet{Giacconi1974}. The nature of the compact object of Cyg X--3 is still ambiguous. 
It can be either a neutron star or a black hole. However, the compact object of 4U 1538--522 is a pulsar, and hence a neutron star.
Observations \citep{{Vilhu2009},{Reynolds1992}} have indicated that the mass accretion mode for these two HMXB systems is via strong stellar winds of the companion stars. While HMXBs are well-studied, the effects of the magnetic field on the ionised spectral lines have not been explored.
Here, we study such effects.
Since a significant amount of data are available and more will be available in near future, our work can also be extended to many other similar HMXBs
to derive their underlying physical conditions.

This paper is organised as follows. In
section 2, we describe our models and calculations. Section 3 describes our findings and the results for the HMXB systems, Cyg X--3 and 4U 1538--522. In section 4, we summarises our results and conclusions. 

\section{Calculations}
In this section we briefly present our numerical calculations which are carried out using the spectral synthesis code CLOUDY (https://nublado.org) which is 
a micro-physics code 
based on a self-consistent $\textit {ab initio}$ calculation of thermal, ionisation, and chemical balance of non-equilibrium gas and dust 
exposed to a source of radiation. It predicts the resultant spectra and vice versa over the entire range of the electromagnetic spectra using a minimum 
number of free parameters. We use an improved version of c-17.02 of CLOUDY which includes 625 species
including atoms, ions and molecules and utilises five
distinct databases: H-like and
He-like iso-electronic sequences \citep{2012Porter}, Stout \citep{Lykins2015}, CHIANTI \citep{2012Landi},
LAMBDA \citep{2005Schoier}, and the H$_{2}$ molecule \citep{2005Shaw} to model spectral lines. 
The highly ionised lines modelled here belong to either H-like or He-like iso-electronic sequences. 
It is to be noted that atoms of the H-like iso-electronic sequence have one bound electron, and
atoms of the He-like iso-electronic sequence have two
bound electrons. CLOUDY uses a unified
model for both the H-like and He-like iso-electronic sequences, that extends from H to Zn, as described by \citet{2012Porter}. 
The new version, c-17.02 of CLOUDY, uses improved energy levels of K${\alpha}$ transitions \citep{Chakraborty2020a}.  
The DR data used in Cloudy are taken mainly from http://amdpp.phys.strath.ac.uk/tamoc/DR/.
A detailed description can be found in \citet{{Ferland2013}, {Ferland2017}, {Chakraborty2020b}, 
{Chakraborty2020c}}.

Previously, we have performed a spectroscopic modelling of highly-ionised X-ray lines from four low-mass X-ray binary (LMXB) systems \citep{Shaw2019} 
using CLOUDY. In those LMXB systems, the mass transfer was via Roche Lobe overflow and 
the observed lines were originated in the associated accretion discs. 
In that work, we estimated various physical parameters like density, radiation field, chemical abundances and 
found that the highly ionised iron lines are 
sensitive to magnetic field. Based on this fact and the observed intensity of the highly ionised iron lines, we estimated an upper limit for the accretion 
disc magnetic field. Here 
our aim is to determine various physical parameters as well as to put an upper limit of the strength of the magnetic field where the highly ionized X-ray 
lines originate, following \citep{Shaw2019}. However, there is some differences in the current modelling. The HMXBs considered here are fed by stellar winds,
whereas the
previously studied LMXBs \citep{Shaw2019} are fed through Roche Lobe overflow. 
 
\subsection{Models} 
\label{subsec:Models}
The companion stars of Cyg X--3 and 4U 1538-522 are a Wolf-Rayet \citep[WR; ][]{Kerkwijk1996} and a B0Iab star \citep{Reynolds1992}, respectively. 
The wind velocity profile 
from a super-giant (SG) star is described by $\beta$-velocity 
law \citep{Castor1975},
\begin {equation}
v(r)=v_{\infty}\times (1-r_{SG}/r)^{\beta}.
\end {equation}
Here $\beta$ is velocity gradient and $v_{\infty}$ denotes the terminal velocity. In general, $v_{\infty}$ ranges from 1000 to 3000 km s$^{-1}$.
For a steady-state wind, density profile n$_H$(r) depends on the mass-loss rate $\dot{\text{M}}$. The hydrogen 
density profile n$_H$(r) is given 
by the equation,
\begin {equation}
n_{H}(r)=\frac{\dot{\text{M}}}{4 \pi \text{r}^2 \\{m_H} \mu \text{v(r)}},
\end {equation}
where $\mu$ and m$_{H}$ represent the mean molecular weight of the gas and mass of hydrogen atom, respectively. For simplicity, we assume $\beta=0$ and a 
constant mass loss rate. As a result, the density decreases with distance in a power-law with a power $-2$,
\begin {equation}
n_{H}(r) \propto{r}^{-2}.
\end {equation}
To be noted here that CLOUDY measures distance from the ionizing source which is fixed in the code. In our models, the compact star is the main source of X-ray ionization. Hence, we use the wind density distribution along the line of sight from
the X-ray source calculated for a wind spherically symmetric with
respect to the compact star. We consider the elemental abundances of the wind in our calculation. Earlier \cite{Szostek2008} also assumed similar density profile for Cyg--X3. Hence, we use the following radius dependent power-law density profile 
\begin {equation}
n_{H}(r)=n_{H}(r_{0})\times (r/r_{0})^{-2}, 
\end {equation}
for all our models presented here. Here $n_{H}(r_{0})$ is the density at the illuminated face at $r_{0}$ and 
the radius $r$ is centered on the compact star.
For hydrogen density $n_{H}(r)$ (cm$^{-3}$), we use the total hydrogen density. In general, physical conditions vary at different phases of the orbit. In our model we incorporate this by using a covering factor, $f$, where $f$ is the 
fraction of 4$\pi $ $\it{sr}$ covered by gas, as viewed from the central source of radiation. The value 
of $f$ lies between 0 to 1. 

In our models presented here, we assume that 
the stellar wind is irradiated by radiation field coming from both the stars of the binary system. A blackbody is considered to explain the X-ray emission 
from and/or near the compact star. As a result, our considered radiation is composed of three components: i) a black body source appropriate 
to the companion stars' surface temperature 
(T$_{BB}$), ii) another blackbody continuum with temperature equivalent to a few tenth of one keV arising from the thermal emission near the compact star's 
surface \citep{Vander2005}, and iii) a Comptonised power-law continuum  $\nu$$^{-\alpha}$. This 
power-law continuum behaves 
as $\nu$$^{5/2}$ at lower energy to account for self-absorbed synchrotron \citep{1979Rybicki}, $\nu$$^{-2}$ at 
higher energies and $\nu$$^{-\alpha}$ between 10 microns and 50 KeV. The photon index, $\alpha$, 
is a free parameter. It is to be noted that, without 
considering the power-law continuum, we cannot predict the observed fluxes of Fe XXV and Fe XXVI lines. 

To study the effect of magnetic field in our models, we use a tangled magnetic field (in units of G) following the equation
\begin {equation}
B = B_{0} \times (n_{H}(r) / n_{H}(r_{0}))^{\gamma /2}.
\end {equation}
where, $B_{0}$ is a free parameter and  the term in parenthesis is the ratio of the  density at a distance $r$ to the density at the 
illuminated face at $r_{0}$. The standard value of the adiabatic index $\gamma$ for a tangled magnetic field is 4/3. Lack of information on 
the angle between the radiation 
field of the central object and the magnetic field restrains us from using an ordered magnetic field.
Here we propose that with {\it a prior} knowledge of other physical conditions, the strengths of highly ionised X-ray 
lines originating in the stellar wind of the companion star 
can be utilised to estimate an upper limit of the local magnetic field.
Earlier, other modellers, e.g. \citet{Szostek2008}, used CLOUDY to study the effect of X-ray on the stellar wind. 
The current updated c-17.02 version of CLOUDY uses improved energy levels 
of K${\alpha}$ transitions \citep{Chakraborty2020a} which 
improves the model predictions. In addition to that, we study the effects of magnetic field on highly ionised X-ray lines.

Cloudy has a built-in optimization process based on phymir algorithm \citep{vanhoof1997} 
which calculates a non-standard 
goodness-of-fit estimator 
$\chi$$^2$ and minimises it by varying input parameters. The $\chi^2$ is determined by the following relation,
\begin{equation}
\chi^2 =\sum_{i=1}^n(\frac{(M_i-O_i)}{min(M_i, O_i)\sigma_i})^2.
\end{equation} Here, $n$ is the number of emission lines used in the model, $O_i$ is the observed value for 
the i-th observable, $M_i$ is the modelled value for this observable, and $\sigma_i$ is the relative error in the observed value.
When the number of observed lines are greater than the number of input parameters, we use this method \citep{{2006Shaw},{2016Shaw}}. Otherwise, we 
perform a grid of models in the parameter space (within an acceptable range) \citep{{Mondal2019},{2018Rawlins}}. We then pick the parameter values which predict 
better match to the observed values and finally fine-tune relevant input parameters so that the observed data can be matched maximally.

\subsection{Effects of input parameters} \label{subsec:parameters}
\begin {figure}
\includegraphics[scale=0.5]{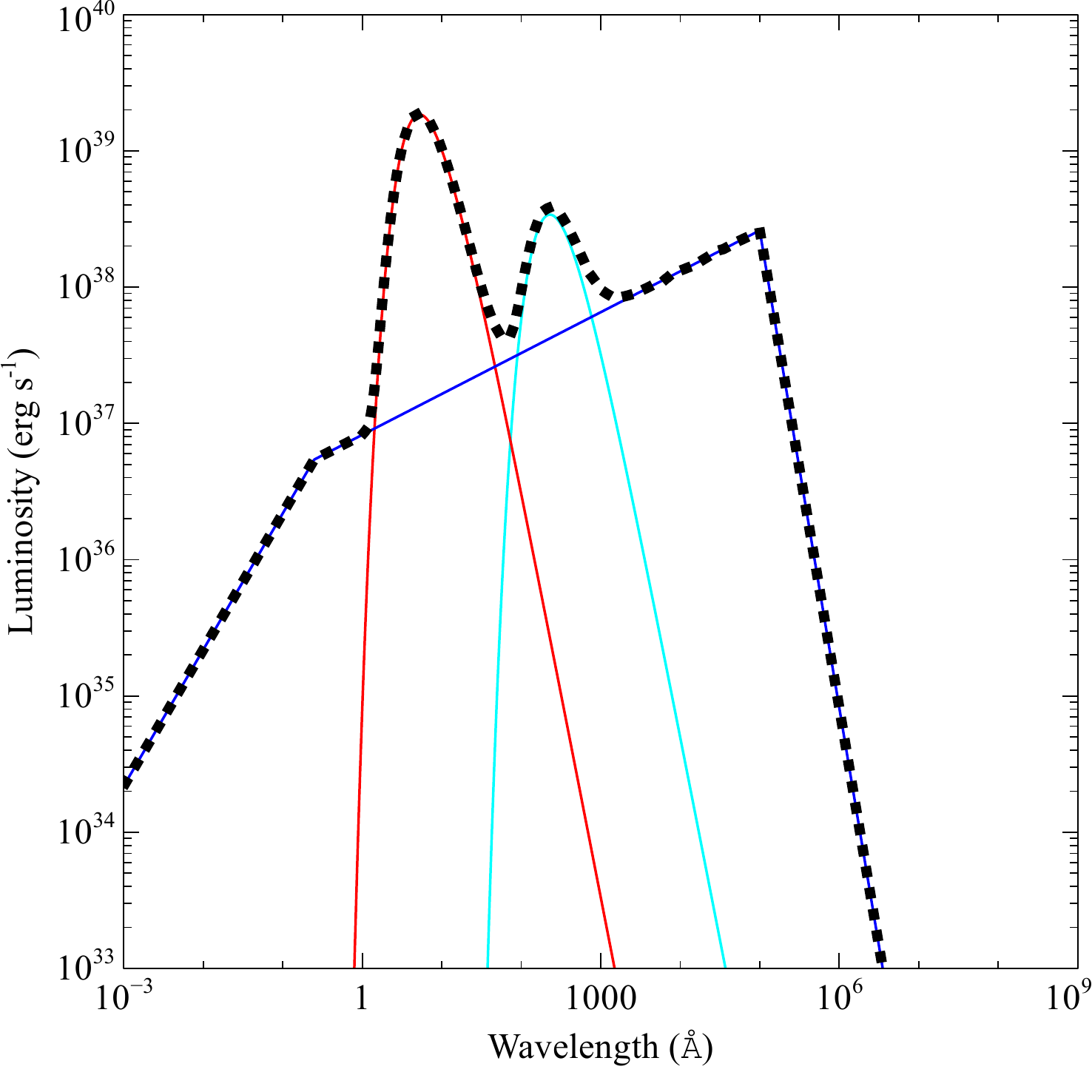}
\caption{The red line
represents incident blackbody continuum  from the thermal emission near the compact star's 
surface. The blue and cyan lines represent Comptonised power-law continuum and a black body continuum appropriate 
to the companion star's surface temperature. The black dots represent the total incident continuum as a function of wavelength at 
the illuminated face of the ionised gas for the sample model. }
\label{fig:fig1}
\end {figure}

\begin {figure}
\includegraphics[scale=0.5]{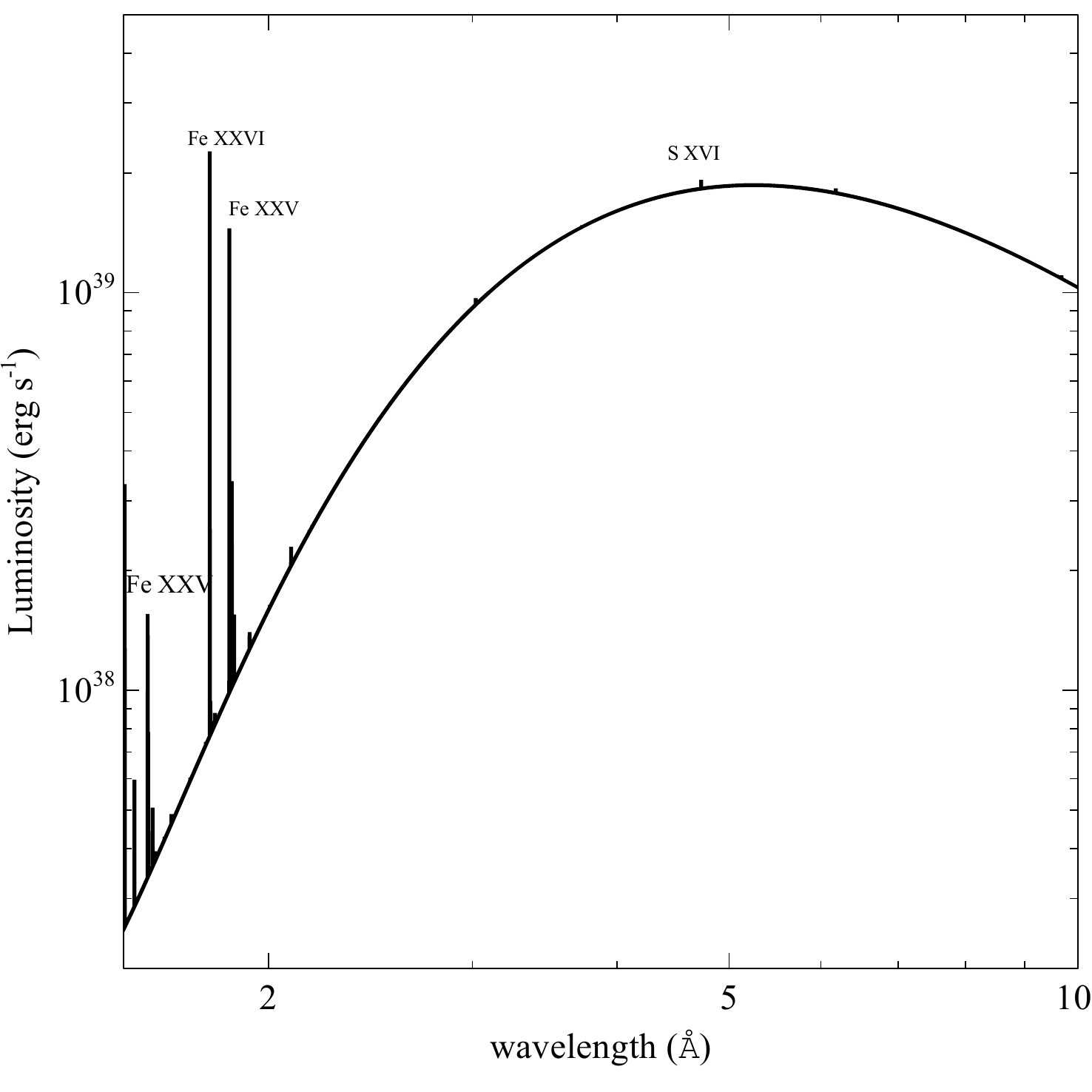}
\caption{Transmitted continuum as a function of wavelength for the sample model. }
\label{fig:fig2}
\end {figure}

Here we consider a sample model for HMXBs and show the effects of various input parameters on Fe XXVI and Fe XXV line strengths 
as both the HMXBs considered here show Fe XXV and Fe XXVI lines. We assume a solar 
metallicity \citep{Grevesse2010} stellar wind irradiated by radiation field coming from both the stars of the binary system consisting of (i) a blackbody source 
at 160,000 K 
with luminosity 
10$^{38.65}$ erg s$^{-1}$ mimicking the radiation from a WR companion star, (ii) another blackbody continuum with temperature 0.7 keV and luminosity 
10$^{39.4}$ erg s$^{-1}$ presenting the 
thermal emission from the compact star, and (iii) a Comptonised power-law continuum with a photon index 1.3 and luminosity 10$^{38.3}$ erg s$^{-1}$. 
The temperature, at which Fe XXV and Fe XXVI lines form, is much higher than the sublimation temperatures of graphite and silicate dust grains. 
Hence we do not include dust in our calculation. Further, we assume $n_{H}(r_{0})$ = 10$^{12.3}$ cm$^{-3}$ and 
the wind velocity to be 1500 km s$^{-1}$. We do not resolve the binary stars to keep the uncertainties of the model low. The inner radius is 10$^{11}$ cm away from the compact star and the extension of the ionized gas is 
set by a given column density of Hydrogen. For the sample model, this value is 10$^{22.5}$ cm$^{-2}$. 
The sample model parameters 
are shown in Table \ref{tab:table 1}. Fig.\ref{fig:fig1} shows the total incident continuum and its constituting individual parts.
Fig.\ref{fig:fig2} shows the transmitted continuum of the sample model as a function of wavelength. The transmitted continuum consists of the attenuated incident continuum and the outward component
of the diffuse continuum and line emission. The resolving power of High Energy Transmission Grating (HETG) instrument on-board $\it{Chandra}$ varies from $\sim$ 800 at 1.5 keV to $\sim$ 200 at 
6 keV. We thus set the coarse continuum mesh resolution E/$\Delta$E = 660. 

\begin{table}
	\centering
	\caption{Input parameters for our sample HMXB model (see section \ref{subsec:parameters}.)}
	\label{tab:table 1}
	\begin{tabular}{lr} 
		\hline
		Physical parameters & Values\\
		\hline
Power law: photon index, luminosity (erg s$^{-1}$) & 1.3, 10$^{38.3}$\\
T$_{BB}$ (companion star), luminosity (erg s$^{-1}$) & 160000 K, 10$^{38.65}$\\
T$_{BB}$ (compact star), luminosity (erg s$^{-1}$) & 0.7 keV, 10$^{39.4}$\\
Density n(r$_{0}$) (cm$^{-3}$) & 10$^{12.3}$\\
Inner radius r$_{0}$ (cm) & 10$^{11}$\\
Wind (km s$^{-1}$) & 1500  \\
Abundance & Solar\\
Hydrogen column density N(H) (cm$^{-2}$) & 10$^{22.5}$\\
\hline
\end{tabular}
\end{table}

The surface temperature of the companion star is much less than the excitation temperatures of 
Fe XXV and Fe XXVI.
Hence, the luminosities of Fe XXV and Fe XXVI lines do not depend on the surface temperature of the companion star. 
So, among all the parameters, the temperature of the blackbody continuum 
from the companion star surface has no effect on the fluxes of Fe XXV and Fe XXVI lines of the sample model.
Therefore, we do not consider the blackbody radiation from the companion star while modelling Cyg X--3 and 4U 1538--522.

To gauge the effects of a parameter on the line fluxes, we  only vary that parameter while keeping all others fixed.
Table \ref{tab:table 2} shows how the line fluxes change with changing parameters.
\begin{table}
	\centering
	\caption{Effect of input parameters on Fe XXV and Fe XXVI line fluxes}
	\label{tab:table 2}
	\begin{tabular}{lllc} 
		\hline
		Parameter & Fe XXVI  & Fe XXV  & Fe XXV/Fe XXVI \\
		& (erg s$^{-1}$)& (erg s$^{-1}$)&\\
		\hline
Default case & 10$^{34.767}$ & 10$^{33.937}$ & 0.148\\
n(r$_{0}$)=10$^{12.6}$ (cm$^{-3}$)& 10$^{34.896}$ & 10$^{34.334}$ & 0.274\\ 
n(r$_{0}$)=10$^{12.0}$ (cm$^{-3}$)& 10$^{34.633}$ & 10$^{33.569}$ & 0.086\\
r$_{0}$=10$^{11.3}$ cm & 10$^{34.942}$ & 10$^{35.077}$& 1.365\\ 
r$_{0}$=10$^{10.7}$ cm & 10$^{33.766}$ & 10$^{31.724}$& 0.009\\
N(H)=10$^{23}$ cm$^{-2}$ & 10$^{35.360}$ & 10$^{34.901}$&0.347\\
N(H)=10$^{22}$ cm$^{-2}$ & 10$^{34.942}$ & 10$^{33.378}$& 0.122\\
Z=2$\times$Z$\odot$ & 10$^{34.991}$ & 10$^{34.234}$ & 0.175\\
Z=0.5$\times$Z$\odot$ & 10$^{34.506}$ & 10$^{33.638}$& 0.135\\
Photon index=1 & 10$^{35.077}$ & 10$^{35.018}$& 0.873\\
Photon index=1.6 & 10$^{34.375}$ & 10$^{31.008}$& 0.004\\
T$_{BB}$= 0.6 keV & 10$^{33.449}$ & 10$^{33.602}$& 0.169\\
T$_{BB}$= 0.8 keV & 10$^{35.025}$ & 10$^{33.988}$& 0.092\\
\hline
\end{tabular}

\end{table}

We notice that the wind velocity has very negligible effect on the Fe XXV and Fe XXVI line fluxes.
To check the effect of changing hydrogen number density, we change the density 
by $\pm$0.3 dex from the 
default value, 10$^{12.3}$ cm$^{-3}$.
It is clear from the plot that the 
line fluxes are density dependent (below the critical density) and Fe XXV line is more sensitive as discussed in the 
introduction. Individual line fluxes as well as Fe XXV / Fe XXVI line ratio increase with increasing density. In the following, we discuss the 
effects of other parameters.

We vary the inner radius by $\pm$0.3 dex from the default value to check its effect. For a given source luminosity, the radiation flux decreases as $r^{-2}$, and hence the temperature decreases with increasing inner radius. The collisional ionisation is temperature dependent, and its value decreases with decreasing temperature. Hence, individual line fluxes as well as Fe XXV / Fe XXVI line ratio increase with increasing inner radius. For the sample model, the effect of inner radius is more profound for Fe XXV.

The extension of the ionised gas is defined by N(H). So, a higher N(H) means a greater extension of the
ionised cloud. As a result, the Fe XXV and Fe XXVI line fluxes increase with increasing N(H). However, Fe XXV line flux increases more than the Fe XXVI line. Hence, Fe XXV/Fe XXVI ratio increases with increasing N(H).

To check the effect of changing metallicity, the abundance of elements heavier than He,
we change the metallicity by a factor of 2 from the default value. The density 
considered in the sample model are lower than the critical densities of Fe XXV and Fe XXVI. Hence, the luminosity of both the lines increases with increased metallicity. In addition to this, metallicity plays an important role in gas cooling. An Increased metallicity decreases the gas temperature and affects physical processes which are temperature dependent. 
As discussed earlier, the collisional ionisation is temperature dependent. Hence, increased 
metallicity increases Fe XXV/Fe XXVI.

To check the effect of changing photon index of the power-law spectrum we run two models with photon index 1.0 and 1.6. The number of incident photon flux ($\propto$ $\nu^{-photon-index}$) increases as photon index decreases, and it is more for smaller frequency. Hence, the line fluxes as well as Fe XXV/Fe XXVI increases as photon index decreases.

To check the effect of the compact star's blackbody temperature, we run two cases with surface temperature 0.8 keV and 0.6 keV.
A blackbody with a few tenth of keV temperature peaks in X-rays. Lowering this temperature reduces the height of the 
peak and shifts the peak to higher wavelengths. Hence, the continuum luminosity is affected by changing the blackbody temperature. The gas becomes less ionised and temperature 
also decreases with lower surface temperature. As a result, luminosity/intensity of the X-ray lines also decreases with decreasing surface temperature. With a higher surface temperature, the Fe XXV/Fe XXVI ratio decreases.  

\section{Results}
\label{sec:Results}
In this section, we present our results on HMXBs Cyg X--3 and 4U 1538--522, and elaborate on main findings. First, we  discuss results of Cyg X--3 in detail. 
Then we discuss the results for 4U 1538--522. 

\subsection{Cyg X--3} 
\label{subsec:Cyg X--3}
Cyg X--3 (20$^h$ 32$^m$ 25.78$^s$, +40$^{\circ}$ 57' 27.9'') \citep{Cutri2003} is a well-observed HMXB in the constellation Cygnus, 7.4 
kpc \citep{McCollough2016} away from us.
The nature of the compact star is not well understood, 
it can either be 
a neutron star or a black hole. Whereas, the companion star of this binary system is a Wolf-Rayet 
star \citep{Kerkwijk1996} with an orbital period of 4.8 hours \citep{Liu2007}.  \citet{Koljonen2017} 
had estimated the 
mass of the WR star to be 
8-14 M$_{\odot}$. Whereas, the mass of the compact object is 2.4$^{+2.1}_{-1.1}$ M$_{\odot}$ \citep{Zdziarski2013}.

For a binary stellar system, the average distance between the two stars with known masses and known orbital period can be calculated using the 
simplistic expression of Kepler's 3rd law, and using that one can calculate the average distance between the compact star and the companion star 
for Cyg X--3 to be $\approx$ 2-3 
$\times$ 10$^{11}$ cm.
Assuming a stellar wind with a velocity of the order of 10$^{3}$ km s$^{-1}$, the Bondi-Hoyle wind accretion radius is 
found to be $\approx$ 3-12 $\times$ 10$^{10}$ cm. 

The compact object is very close to the Wolf-Rayet star in this case. Hence, the materials of the stellar wind get pulled away from the 
Wolf-Rayet star by the compact object and become hotter and shine in X-rays. This system exhibits many interesting properties like 
jets \citep{Dubus2010} and radio flare associated 
with $\gamma$ ray emission etc \citep{{Corbel2012},{Pahari2018}}. But here we focus only on the spectroscopic study of highly ionised X-ray 
lines.

Spectroscopic observation of Cyg X--3 has been done by many groups \citep{{Fender1999}, {Paerels2000}, {Vilhu2009}, {Kallman2019}}. \citet{Fender1999} performed 
IR observations, whereas
\citet{Paerels2000} and \citet{Vilhu2009} used  {\textit{Chandra}} to study Cyg X--3 in X-rays. They detected Si XIV (2.005 keV), S XVI (2.62 keV), Fe XXV (6.7 keV) 
and Fe XXVI (6.966 keV) lines 
using {\textit{Chandra}}. In this work, we model the X-ray line intensities 
observed by \citet{Vilhu2009}
as they provided the line flux values in a tabular form which eases modelling. \citet{Vilhu2009} used HETG instrument on-board $\it{Chandra}$. The spectral resolutions are
0.023 Å and 0.012 Å for Medium energy Grating (MEG) and High energy Grating (HEG), respectively. The observed
Fe XXV line is a composite of 1.869 {\AA}, 
 1.859{\AA}, 1.852 {\AA} lines arising partially
in the wind with intensity ratios 0.35:0.45:0.19,  and partially in
the region where the Fe XXVI line originates. Earlier \citet{Szostek2008} used CLOUDY (v6.02) to study the effect of stellar 
wind on the X-ray lines for the same source observed using BeppoSAX.  
Our current model uses {\textit{Chandra}} data with higher resolution and we use a more advanced version of Cloudy, c-17-02, 
which is capable to predict 
the line energies and line intensities of highly excited ions more accurately \citep{Chakraborty2020a}. 
In addition, we also study the effect of 
magnetic field on highly ionised lines. 

\begin {figure}
\includegraphics[scale=0.5]{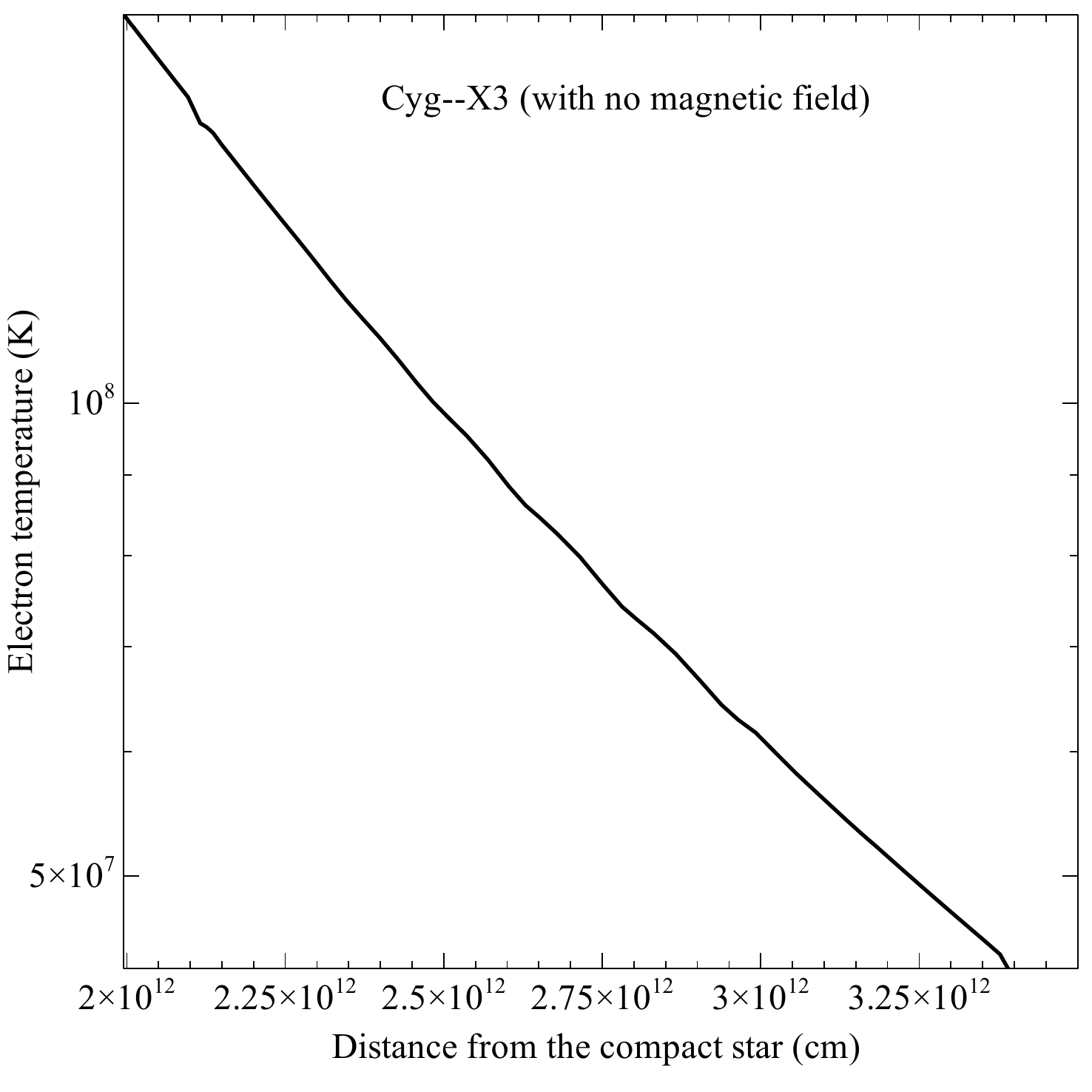}
\caption{The variation of electron temperature as a function of distance from the compact star.}
\label{fig:cygx3_temp_nomag}
\end {figure}

\begin {figure}
\includegraphics[scale=0.5]{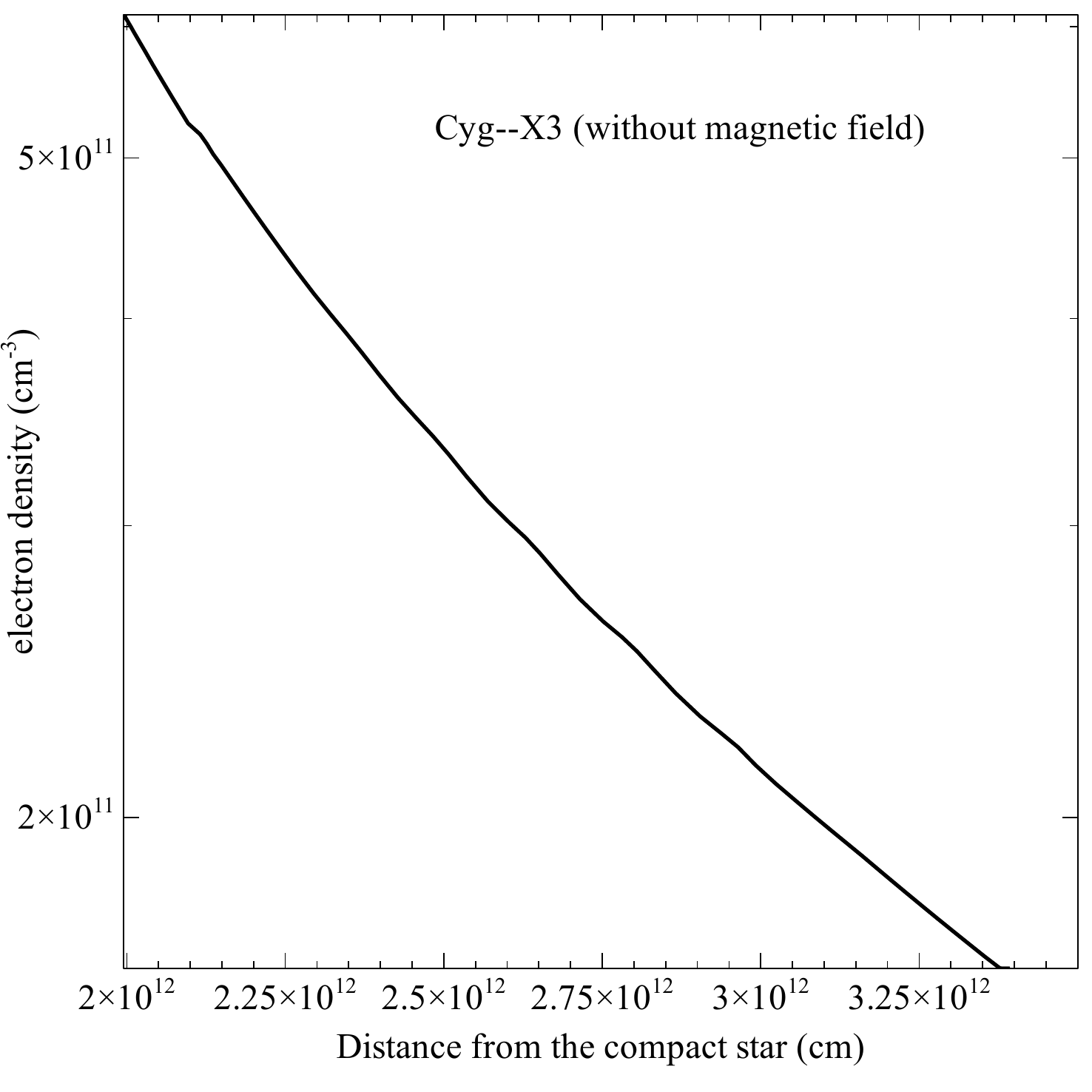}
\caption{The variation of electron density as a function of distance from the compact star.}
\label{fig:cyg_nomag_eden}
\end {figure}

\begin {figure}
\includegraphics[scale=0.5]{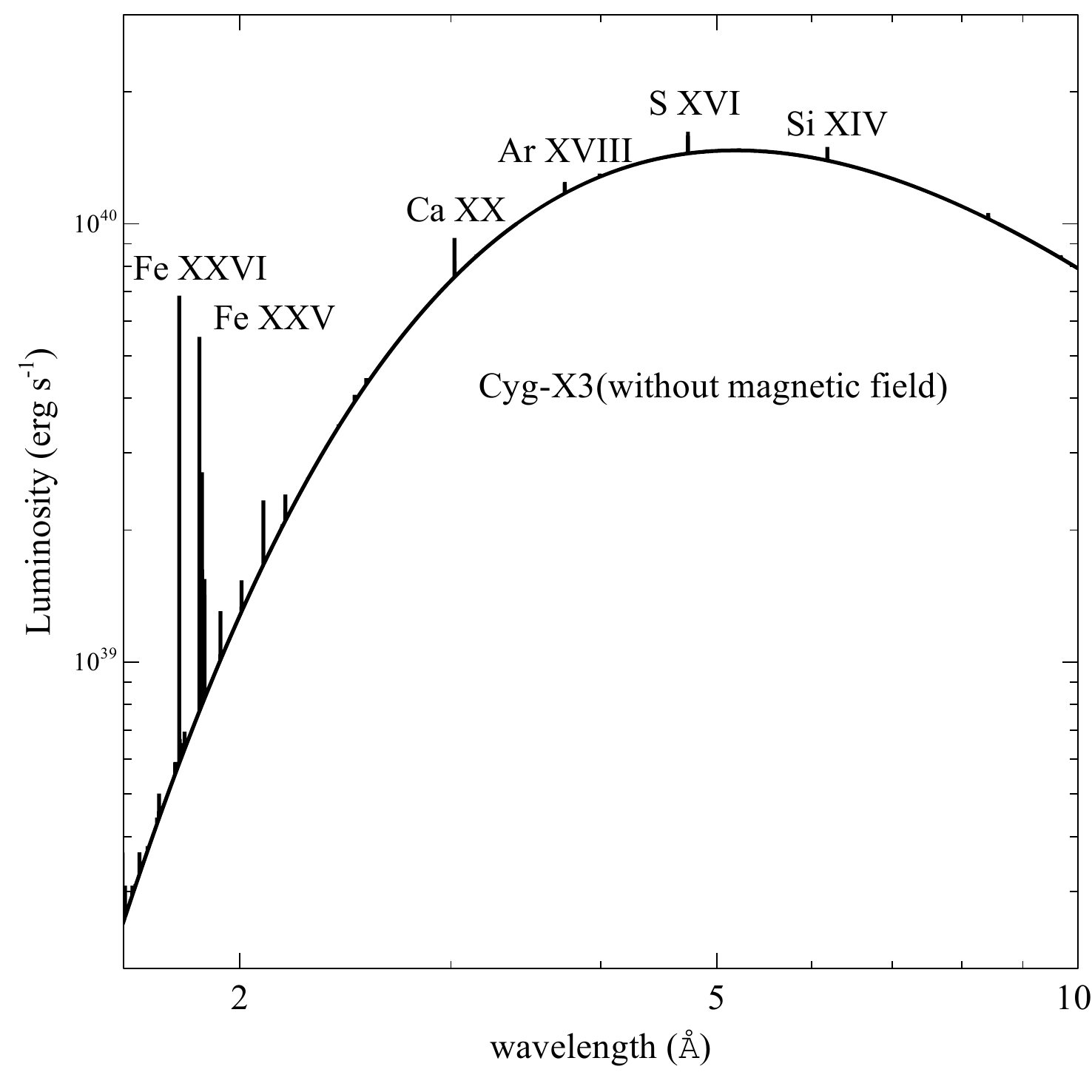}
\caption{Predicted spectra for Cyg--X3 (without magnetic field).}
\label{fig:cyg_nomag_spectra}
\end {figure}

In our model, we assume that the stellar wind is irradiated by the compact object with a blackbody temperatures 0.7 KeV.
The corresponding  
luminosity is 10$^{39.40}$ erg s$^{-1}$ \citep{Vilhu2009}. In addition to this, there is a Comptonised continuum having a power-law 
with the photon index 
$\alpha$. It is to be noted that the blackbody radiation from the companion star does not effect the 
X-ray line intensities. However, the metallicities of the companion star is important.
We consider the typical 
Wolf-Rayet abundances assuming a WN(4-7) star: H=0.1, He=10, C=0.56, N=40, O=0.27 \citep{Vilhu2009}. The abundances of other elements are fixed at their solar values. 
In general, the stellar wind velocity of HMXBs range from 1000 to 3000 km s$^{-1}$ \citep{Kaper1995}. In this model, we choose a stellar wind with velocity 
2000 km s$^{-1}$. 
\citet{Vilhu2009} noticed that the line intensities are maximum around phase 0.5, when the compact star is in front. Here the geometry is complex, and as a first approximation we fix the value of the covering factor to be 0.25 for our simple model.
From earlier literature we find that the value of photon index for this source is around 2 for soft X-ray observations \citep{Koljonen2018}. We choose 2 as the value of photon index for our model. We fix the elemental abundances with respect to hydrogen (in log scale) as, 
He/H = -0.071, N/H = -2.568, C/H = -3.822, O/H = -3.878, Fe/H = -4.500, S/H = -4.733, Si/H = -4.448, similar to a WN(4-7) star.
\citet{Vilhu2009} performed an 
XSTAR simulation \citep{Bautista2001} and 
predicted the hydrogen density to be $10^{11}$ cm$^{-3}$ and photon index to be 2. Our predicted hydrogen density at the illuminated face is close to this value. Our predicted hydrogen column density also matches to 
the value derived by \citet{Vilhu2009}.
The electron density (Fig.~\ref{fig:cyg_nomag_eden}) and the electron temperature (Fig.~\ref{fig:cygx3_temp_nomag}) vary across the ionized gas. The value of electron temperature averaged over the extension of the cloud is 9.05$\times$10$^{7}$ K. Fig.~\ref{fig:cyg_nomag_spectra} show the predicted transmitted spectra. 
In Table \ref{tab:table 3} we list the model parameters without considering any magnetic field.

\begin{table}
	\centering
	\caption{Predicted model parameters of HMXB Cyg X--3 (without considering magnetic field)
	using CLOUDY (see section \ref{subsec:Cyg X--3} for details).}
	\label{tab:table 3}
	\begin{tabular}{lr} 
		\hline
		Physical parameters &  Predicted values\\
		\hline
		Power law: luminosity (erg s$^{-1}$) &  10$^{38.3}$\\
                T$_{BB}$ (compact star), luminosity (erg s$^{-1}$) & 0.7 keV, 10$^{40.3}$ \\
		Density $n_{H}(r_{0})$(cm$^{-3}$) & $10^{11.35}$\\
		Wind (km s$^{-1}$) & 2000 \\
		Hydrogen column density (cm$^{-2}$) & 10$^{23.2}$\\
		Inner radius r$_{0}$ (cm) & 10$^{12.25}$ \\
		\hline				
	\end{tabular}
\end{table}
 
In Table \ref{tab:table 4} we compare the observed and predicted luminosity of HMXB Cyg X--3. Our predicted luminosities for lines Fe XXVI, Si XIV, S XVI matches very well with the observation. Our predicted total luminosity for Fe XXV also matches with the observation, though the ratios among the Fe XXV lines do not match. It is to be noted that the ionising wind model of \citet{Vilhu2009} could not reproduce the observed line fluxes of Fe XXV and Fe XXVI. In addition to the observed lines mentioned in \citet{Vilhu2009}, we also predict Ca XX and Ar XVIII with luminosity above 10$^{34.5}$ erg s$^{-1}$. These lines were not reported in \citet{Vilhu2009}. This can be due to interstellar absorption and/or subsolar abundances of Ca and Ar. 
To check this, we run a model with Ca and Ar abundances lowered by a factor of -1. and -0.7 (in log scale), respectively. The predicted luminosities decrease by a factor of 10 and 5, respectively. Hence, we conclude that either the abundances of these two elements are subsolar ($\sim$ 1 dex lower) and/or there is absorption by interstellar medium.

Next, we include magnetic field in our model to study its effect on line intensities. Very hot ionised gas cools via cyclotron radiation, and 
we find that with a strength of the magnetic field $\geq$ 10$^{3}$ G, all
the Fe XXV and Fe XXVI line fluxes exceed the observed ranges. At this point, the cyclotron cooling becomes the dominant cooling mechanism.
\begin{table}
	\centering
	\caption{Comparison of observed and predicted line luminosity (in units of 10$^{35}$ erg s$^{-1}$) of HMXB Cyg X--3
          using CLOUDY (see section \ref{subsec:Cyg X--3} for details).}
	\label{tab:table 4}
	\begin{tabular}{llllll}
\hline
	Observed & $\lambda$ & Observed &  Predicted & $\lambda$& Predicted \\          
		Lines & ({\AA}) &Lum & lines & ({\AA}) & Lum \\ 
		\hline
		Si XIV & 6.185 & 1.00$\pm$0.08 & Si XIV & 6.18223 & 1.00\\
		S XVI & 4.733 & 1.80$\pm$0.10 & S XVI & 4.72915 & 1.71\\
		Fe XXV & 1.859 & 3.38$\pm0.60$ & Fe XXV & 1.85951 &1.87\\
		Fe XXV & 1.869 & 2.62$\pm0.60$ & FeXXV & 1.86819 & 0.70\\
		Fe XXV & 1.852 & 1.42$\pm0.60$ & Fe XXV & 1.85040& 4.60\\
		Fe XXVI & 1.780 & 6.05$\pm$0.50 & Fe XXVI & 1.77982 & 6.23\\
		& & & Ca XX & 3.02028 & 1.69\\
		& & & Ar XVIII & 3.73290 & 0.72\\
		\hline
	\end{tabular}
        \end{table}

So, we reoptimise all the parameters including magnetic field. Table \ref{tab:table 5} and Table \ref{tab:table 6} list the predicted model parameters and the line luminosities, respectively. The variation of electron density and electron temperature have features similar to the previous case
without magnetic field. However, the electron temperature averaged over the 
extension of the cloud is 8.66$\times$10$^{7}$ K, less than the previous case without magnetic field. Our model predicts a magnetic field with strength 10$^{2.5}$ Gauss. The predicted spectra is similar to that of without magnetic field.

\begin{table}
	\centering
	\caption{Predicted model parameters of HMXB Cyg X--3 (with magnetic field)
	using CLOUDY (see section \ref{subsec:Cyg X--3} for details).}
	\label{tab:table 5}
	\begin{tabular}{lr} 
		\hline
		Physical parameters & Predicted values\\
		\hline
		Power law: luminosity (erg s$^{-1}$) &  10$^{38.3}$\\
                T$_{BB}$ (compact star), luminosity (erg s$^{-1}$) & 0.7 keV, 10$^{40.3}$ \\
		Density $n_{H}(r_{0})$(cm$^{-3}$) & $10^{11.36}$\\
		Wind (km s$^{-1}$) & 2000 \\
		Hydrogen column density (cm$^{-2}$) & 10$^{23.03}$\\
		Inner radius r$_{0}$ (cm) & 10$^{12.3}$ \\
		Magnetic field Gauss) & 10$^{2.5}$ \\
		\hline				
	\end{tabular}
\end{table}

\begin{table}
	\centering
	\caption{Comparison of observed and predicted line luminosity (in units of 10$^{35}$ erg s$^{-1}$) of HMXB Cyg X--3 (with magnetic field) using CLOUDY (see section \ref{subsec:Cyg X--3} for details).}
	\label{tab:table 6}
	\begin{tabular}{llllll}
\hline
	Observed & $\lambda$ & Observed &  Predicted & $\lambda$& Predicted \\          
		Lines & ({\AA}) & Lum & lines & ({\AA}) & Lum \\ 
		\hline
		Si XIV & 6.185& 1.00$\pm$0.08 &Si XIV &6.18223& 1.02\\
		S XVI & 4.733& 1.80$\pm$0.10 & S XVI& 4.72915 & 1.71\\
		Fe XXV & 1.859 & 3.38$\pm0.60$ & Fe XXV & 1.85951 & 1.87\\
		Fe XXV &1.869 &2.62$\pm0.60$ & FeXXV & 1.86819 & 0.78\\
		Fe XXV &1.852 &1.42$\pm0.60$ & Fe XXV & 1.85040& 4.81\\
		Fe XXVI & 1.780& 6.05$\pm$0.50 & Fe XXVI & 1.77982 & 5.69\\
		& & & Ca XX& 3.02028& 1.65\\
		& & & Ar XVIII & 3.73290& 0.71\\
		\hline
	\end{tabular}
        \end{table}
Previously \citet{2016Hubrig} had observed several Wolf-Rayet stars using FORS2 spectropolarimetry and estimated the highest value for 
the average longitudinal magnetic field to be 327$\pm$141 G. However, \citet{2014Chevroti} have concluded 0 < B$_{wind}$ $\leq$ 1700 G based on 
spectropolarimetric study of 
a sample of 11 bright WR stars. Our predicted  magnetic field ($\approx$ 10$^{2.5}$ G) is consistent with the observations 
of \citet{2016Hubrig}.

\subsection{4U 1538--522}
\label{subsec:4U 1538--522}
The HMXB 4U 1538--522 is an interesting eclipsing HMXB. It was first observed using UHURU satellite by \citet{Giacconi1974} and later followed by 
many observers. It is located at a distance of 
6.4 kpc \citep{{Reynolds1992}, {Clark2004}, {Bailer2018}} 
with an orbital period of 3.73 days \citep{{Clark2000}, {Baykal2006}, {Mukherjee2006}, {Falanga2015}}. This HMXB system is powered by the strong stellar wind 
from a B0Iab star QV Nor \citep{Reynolds1992}, and highly 
ionised Fe XXV and Fe XXVI lines are observed in emission \citep{{Aftab2019},{Mukherjee2006}, {Rodes2011}}. Unlike the previous case, here the compact star is 
a pulsar.
\citet{Reynolds1992} estimated the 
mass of the companion star to be 
19.9 $\pm$3.4 M$_{\odot}$, and the mass of the compact object is 1.06 $\pm$ 0.27 M$_{\odot}$ \citep{Falanga2015}.
\citet{Clark2000} and few more observers suggest that the orbit is probably eccentric, with an eccentricity approximately 0.18. This source 
shows many interesting properties, like significant evolution in the orbital period and cyclotron lines \citep{Hemphill2019} but as stated earlier, our main 
interest is to do spectroscopic study of the highly ionised X-ray lines.

In this work, we study the 
physical properties of HMXB 4U 1538--522 by modelling its X-ray 
line luminosities observed by \citet{Aftab2019} using XMM- {\textit{Newton}}. The line fluxes of these lines were given in units of 
10$^{-4}$ photons cm$^{-2}$ s$^{-1}$ for the eclipse phase. For calculation purpose, we multiply it with 
corresponding line centre energy and convert into ergs cm$^{-2}$ s$^{-1}$ unit. Here we consider $f=0.6$ for eclipse position. Furthermore, we use 0.2 as the photon index as estimated by \citet{Aftab2019} for the eclipse position.

Similar to the earlier system (Cyg X--3),  we assume that the stellar wind is irradiated by the compact object with a blackbody  
temperatures of a few tenth of KeV and a power-law continuum. The blackbody radiation from the companion star has no effect on the X-ray lines, as expected. 
We consider typical 
solar abundances for this source \citep{Grevesse2010}. Here also we do not consider dust grains, as the 
temperature of the region where Fe XXV and 
Fe XXVI lines form is much higher than the dust grain sublimation temperature.
In this model, we choose a stellar wind with velocity 
1500 km s$^{-1}$ \citep{Abbott1982}. In Table \ref{tab:table 7} we list our predicted model parameters. Our predicted hydrogen density at the illuminated face of the wind is 
$10^{11.99}$ cm$^{-3}$ and the illuminated face is 10$^{10.43}$ cm away from the compact star. 

\begin{table}
	\centering
	\caption{Predicted model parameters of HMXB 4U 1538--522 using CLOUDY (without magnetic field).}
	\label{tab:table 7}
	\begin{tabular}{lr} 
		\hline
		Physical parameters & Predicted values\\
		\hline
		Power law: Luminosity (erg s$^{-1}$) & 36.74 \\
		T$_{BB}$ (compact star), luminosity (erg s$^{-1}$) & 0.7 keV, 10$^{38.1}$ \\
		Density $n_{H}(r_{0})$($cm^{-3}$) & $10^{11.99}$\\
		Wind (km s$^{-1}$) & 1500  \\
		Abundance & Solar\\
		Inner radius (r$_{0}$ (cm) & 10$^{10.43}$ \\
		Hydrogen column density ($cm^{-2}$) & $10^{22.25}$\\
		\hline				
	\end{tabular}
\end{table}

\begin{table}
	\centering
	\caption{Comparison of observed and predicted line fluxes (in units of 10$^{-13}$ erg cm$^{-2}$ s$^{-1}$)of HMXB 4U 1538--522 
	using CLOUDY (see section \ref{subsec:4U 1538--522} for details).}
	\label{tab:table 8}
	\begin{tabular}{lllll} 
		\hline
	Observed & Observed &  Predicted & $\lambda$&Predicted \\          
	Lines  &Flux & lines &{\AA}& Flux \\ 
		\hline
		Fe XXV$^a$&  2.04--1.62 & Fe XXV & 1.85040 & 1.91\\
	    Fe XXVI $^b$& 1.55--1.35 & Fe XXVI & 1.77982 & 1.55\\
		\hline
		&$^a$6.700 keV&$^b$6.966 keV \\
	\end{tabular}
        \end{table}

\begin {figure}
\includegraphics[scale=0.5]{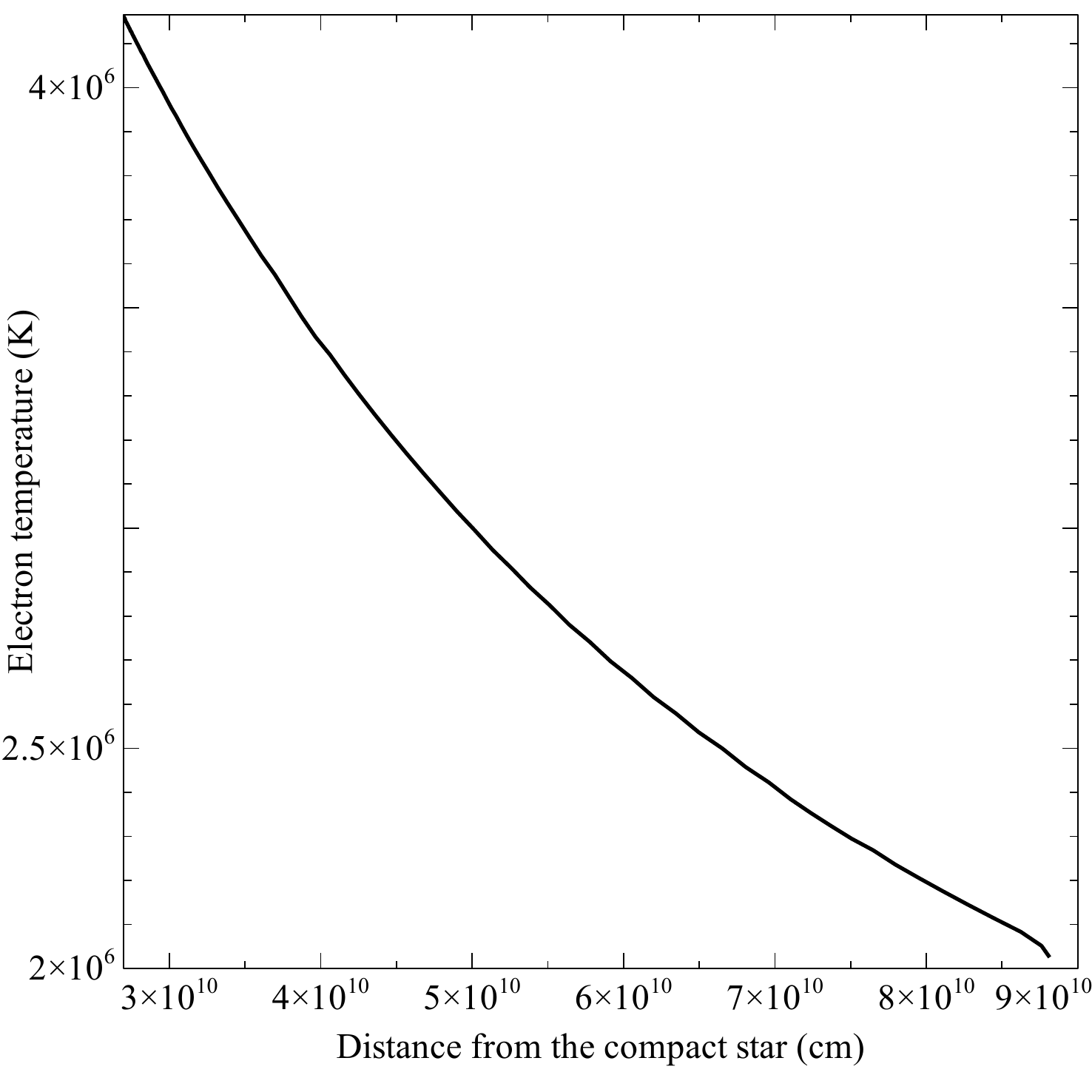}
\caption{The variation of electron temperature as a function of distance from the compact star.}
\label{fig:4U_nomag_temp}
\end {figure}

\begin {figure}
\includegraphics[scale=0.5]{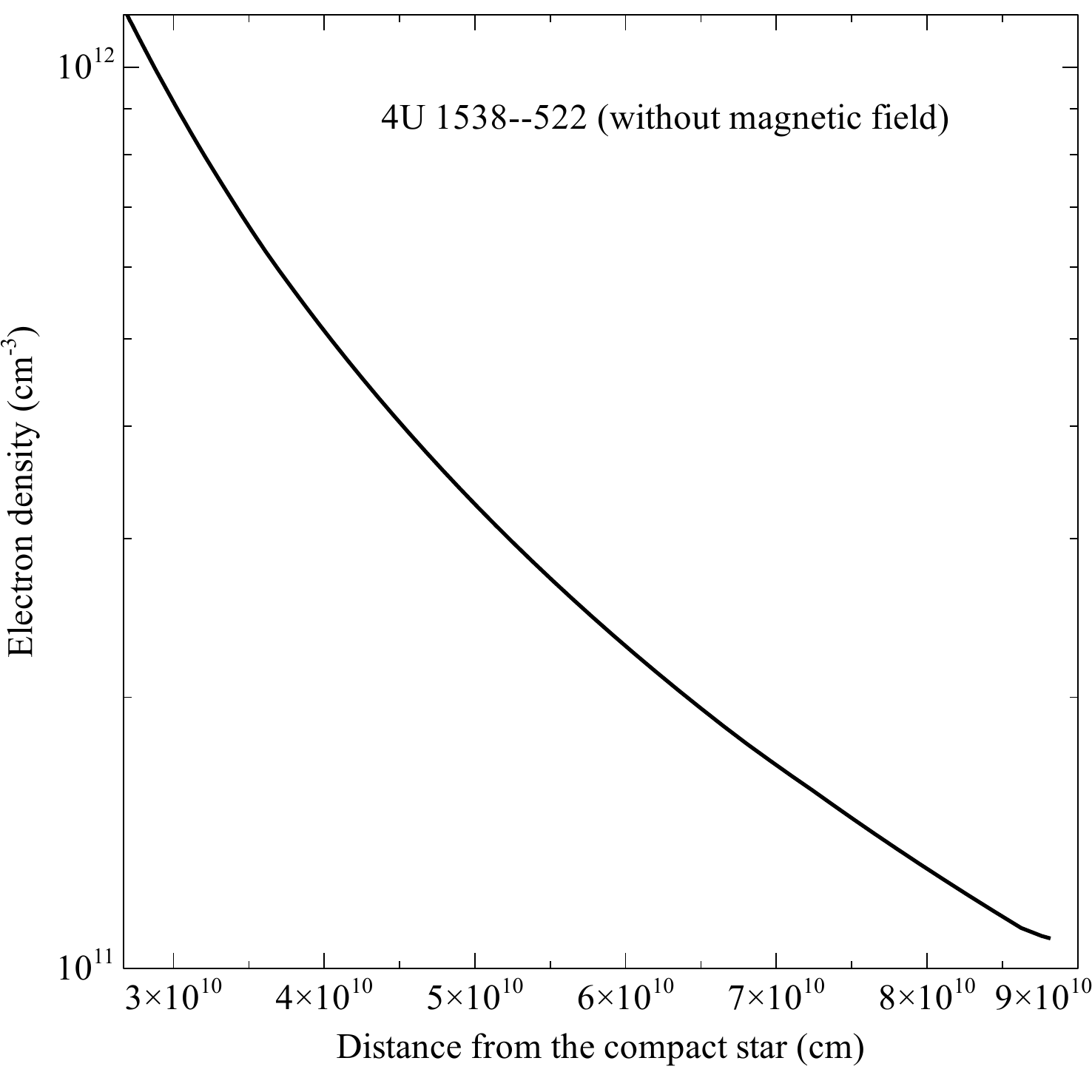}
\caption{The variation of electron density as a function of distance from the compact star.}
\label{fig:4U_nomag_eden}
\end {figure}

\begin {figure}
\includegraphics[scale=0.5]{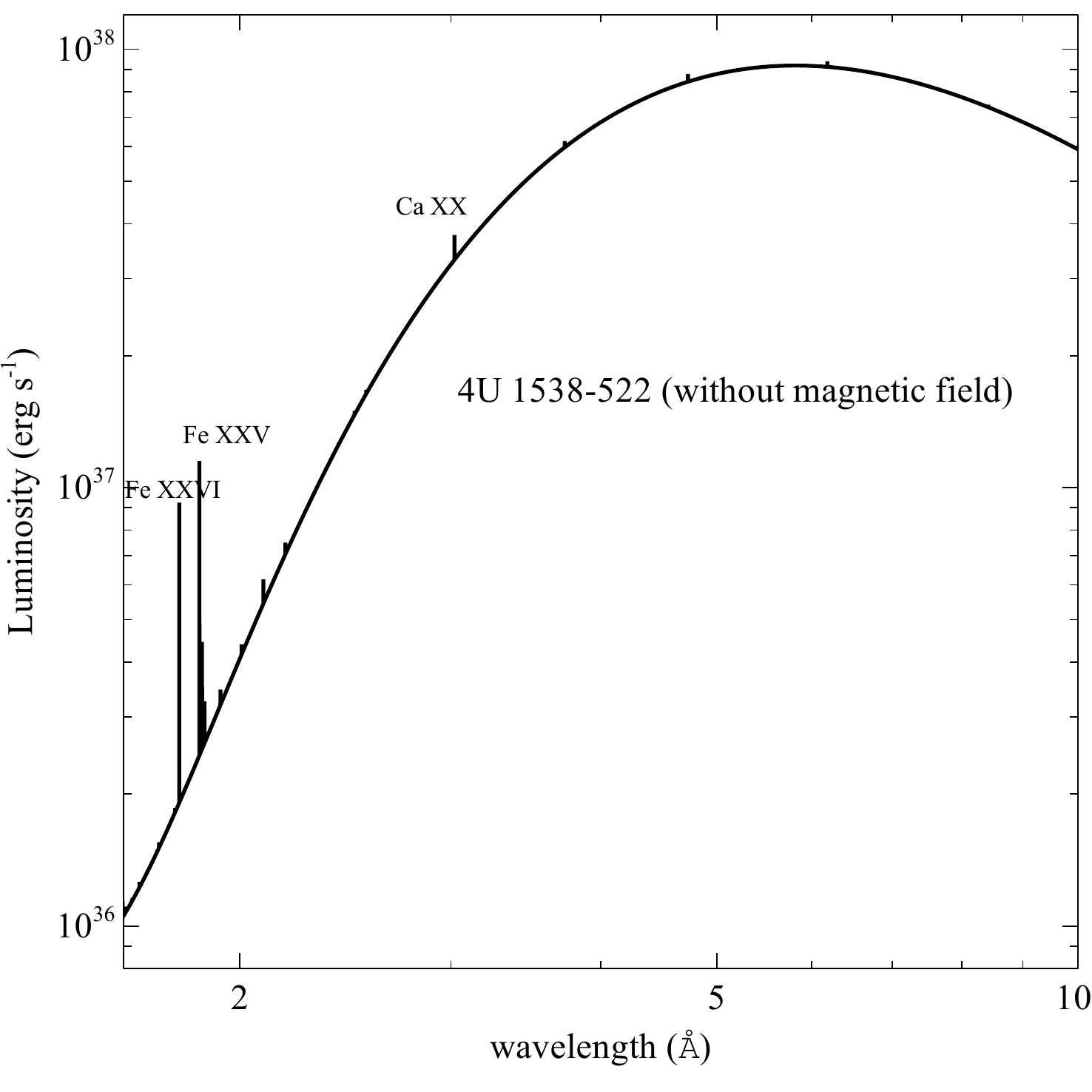}
\caption{Predicted spectra for 4U 1538--422 (without magnetic field).}
\label{fig:4U_nomag_spectra}
\end {figure}

Table \ref{tab:table 8} 
compares the predicted and observed fluxes for eclipse phase of HMXB 4U 1538--522. 
We predict
observed fluxes of both the Fe XXV and Fe XXVI lines simultaneously within the observed range. 
It is to be noted that the model 
parameters could be constrained better for this source if we have more observed lines of different species. 
The electron density (Fig.~\ref{fig:4U_nomag_eden}) and the electron temperature 
(Fig.~\ref{fig:4U_nomag_temp}) vary across the ionized gas. The value of electron temperature averaged over the extension of the cloud is 2.865$\times$10$^{6}$ K. 
Fig.~\ref{fig:4U_nomag_spectra} show the predicted transmitted spectra without any magnetic field. 
Like the previous source, in the final stage, we switch-on the magnetic field in our model to check its effects on the X-ray lines. We find that above a magnetic field $>$ 10$^{3.5}$G, the Fe XXVI and Fe XXVI line fluxes exceed the observed ranges. 
We believe this is caused due to cyclotron cooling. Next, like the earlier source, we reoptimise all the parameters including magnetic field. The predicted electron density, electron temperature, and the predicted spectra are similar to the case without magnetic field. Table \ref{tab:table 9} lists the predicted model parameters. The predicted magnetic field is 10$^{2.5}$ Gauss. Table \ref{tab:table 10} compares the predicted and observed fluxes for eclipse phase of HMXB 4U 1538--522. 
Many groups 
estimated magnetic fields for OB stars \citep{{2016Przybilla},{2017Castro}} and 
found it to be of order of 1kG. 
  
\begin{table}
	\centering
	\caption{Predicted model parameters of HMXB 4U 1538--522 using CLOUDY (with magnetic field).}
	\label{tab:table 9}
	\begin{tabular}{lr} 
		\hline
		Physical parameters & Predicted values\\
		\hline
		Power law: Luminosity (erg s$^{-1}$) & 36.74 \\
		T$_{BB}$ (compact star), luminosity (erg s$^{-1}$) & 0.7 keV, 10$^{38.24}$ \\
		Density $n_{H}(r_{0})$($cm^{-3}$) & $10^{11.99}$\\
		Wind (km s$^{-1}$) & 1500  \\
		Abundance & Solar\\
		Inner radius (r$_{0}$ (cm) & 10$^{10.43}$ \\
		Hydrogen column density ($cm^{-2}$) & $10^{22.05}$\\
		Magnetic field (G) & 10$^{2.5}$\\
		\hline				
	\end{tabular}
\end{table}

\begin{table}
	\centering
	\caption{Comparison of observed and predicted line fluxes (in units of 10$^{-13}$ erg cm$^{-2}$ s$^{-1}$)of HMXB 4U 1538--522 
	using CLOUDY (with magnetic field)}
	\label{tab:table 10}
	\begin{tabular}{lllll} 
		\hline
	Observed & Observed &  Predicted & $\lambda$&Predicted \\          
	Lines  &Flux & lines &{\AA}& Flux \\ 
		\hline
		Fe XXV$^a$&  2.04--1.62 & Fe XXV & 1.85040 & 1.78\\
	    Fe XXVI $^b$& 1.55--1.35 & Fe XXVI & 1.77982 & 1.51\\
		\hline
		&$^a$6.700 keV&$^b$6.966 keV \\
	\end{tabular}
        \end{table}
        

        
\section{Summary and Conclusions}
The HMXB systems are usual sources of strong X-ray lines. To
understand these systems comprehensively, it is crucial to do a
detail {\it ab initio} spectroscopic modeling of these X-ray lines. In
this work, we carry out such a study using the spectral synthesis code
CLOUDY, by self-consistently modelling the observed X-ray line
intensities of Cyg X--3 \citep{Vilhu2009} and 4U 1538--522
\citep{Aftab2019}. We assume these systems as highly-ionised gaseous stellar
winds irradiated by a blackbody continuum with temperature equivalent to a few tenths of one keV of 
the thermal emission from the compact star, and a comptonised power-law continuum  $\nu$$^{-\alpha}$ with photon index $\alpha$.
The photon indices of the
incident radiation of our  models for Cyg X--3 and
4U 1538--522 are fixed to the observed values of 2 and 0.2, respectively. 
It is to be noted 
that the blackbody radiation arising from the companion star does not affect the highly ionised X-ray lines due to its low surface temperature. The presence of magnetic field cools high-temperature ionised gas through cyclotron emission and takes the dominant part in thermal 
cooling. Using the observed line intensities of the Fe XXV and Fe
XXVI lines, we prescribe a plausible way to estimate an upper limit of
the magnetic field of these two systems. For each source we had two models, one without magnetic field and the other one with a tangled magnetic field. Most of the parameters, except the hydrogen column density, have similar values for the cases with and without magnetic field.
Our models predict most of the line fluxes of these two systems within the
observed ranges. We find that the total hydrogen densities are 10$^{11.35}$ or 10$^{11.36}$ cm$^{-3}$ and 10$^{11.99}$ cm$^{-3}$ at the 
illuminated face of the winds for Cyg X--3 and 4U 1538--522, respectively.
The predicted hydrogen
column density for Cyg X--3 and 4U 1538--522 are $10^{23.2}$ cm$^{-2}$
and $10^{22.25}$ cm$^{-2}$ without magnetic field, respectively. The respective values with a tangled magnetic field are, 
$10^{23.3}$ cm$^{-2}$ and $10^{22.05}$ cm$^{-2}$. The value of inner radius with and without a magnetic field is same for 
4U 1538-522 (10$^{10.43}$ cm). However, the value of inner radius for Cyg--X3 is 10$^{12.3}$ and 10$^{12.03}$ cm with and without magnetic field, respectively. We find the most probable strength of magnetic field for 
Cyg X--3 and 4U 1538--522 is $\sim$ 10$^{2.5}$ G. Future polarisation measurements may verify our predictions. 
Finally, we would like to point out that the models adopted here can be further improved in future. 
With a larger number of observed lines at different phases, it would definitely be possible to explore more complex geometry and clumpiness in the wind using, pyCloudy, a pseudo-3D model for CLOUDY.

\section{Acknowledgements}
Gargi Shaw acknowledges WOS-A grant (SR/WOS-A/PM-9/2017) from the Department of Science and Technology, India, and also like to thank the 
Department of Astronomy and Astrophysics, TIFR for its support. We thank the anonymous referee for his/her thoughtful suggestions.

\section{Data Availability}

Simulations in this paper made use of the code CLOUDY (c17.02), which can be down-loaded freely at https://www.nublado.org/. 
The model generated data are available on request.











\bsp	
\label{lastpage}
\end{document}